\documentstyle[aps,twocolumn,epsfig]{revtex}
\begin{document}
\newcommand{\ve}[1]{\mbox{\boldmath $#1$}}
\twocolumn[\hsize\textwidth\columnwidth\hsize
\csname@twocolumnfalse%
\endcsname
 
\draft

\title {Analytical results for a trapped, weakly-interacting 
        Bose-Einstein condensate under rotation}
\author{A. D. Jackson$^1$ and G. M. Kavoulakis$^2$} 
\date{\today} 
\address{$^1$Niels Bohr Institute, Blegdamsvej 17, DK-2100 Copenhagen \O,
        Denmark, \\
         $^2$NORDITA, Blegdamsvej 17, DK-2100 Copenhagen \O, Denmark} 
\maketitle
 
\begin{abstract}

We examine the problem of a repulsive, weakly-interacting 
and harmonically trapped Bose-Einstein condensate under 
rotation. We derive a simple analytic expression for the 
energy incorporating the interactions when the angular 
momentum per particle is between zero and one and find  
that the interaction energy decreases linearly as a function 
of the angular momentum in agreement with previous
numerical and limiting analytical studies. 

\end{abstract}
 
\vskip0.5pc]

One of the questions that followed the experimental realization
of Bose-Einstein condensation \cite{GSS} in vapours of alkali atoms 
\cite{RMP} is how they behave under rotation.  The experimental 
observation of vortex states in a two-component system has
been reported by Matthews {\it et al.} \cite{JILA}, and the
existence of vortex states in a stirred one-component Bose-Einstein 
condensate has been demonstrated by Madison {\it et al.} \cite{Madison}.
Theoretical studies have also addressed this problem both in
the Thomas-Fermi limit of strong interactions between the atoms 
\cite{Rokhsarv,Fetter,Feder} and in the limit of weak interactions
\cite{Wilkin,Rokhsar,Ben,Bertsch,KMP,JKMR}. 

In the limit of weak interactions and for a given angular momentum $L$, 
there is a large degeneracy corresponding to the ways of distributing $L$ 
units of angular momentum among $N$ harmonically confined atoms.  
Diagonalizing the interaction within this space of degenerate states, 
Bertsch and Papenbrock \cite{Bertsch} determined the ground state energy 
numerically for a limited number of particles and an effective repulsive
interaction between them; the same method can 
evidently provide the complete excitation spectrum of the system.  One 
remarkable observation is that, for $2 \le L \le N$, the ground-state energy 
appears to have a particularly simple form with an interaction energy 
which decreases linearly with $L$.  This result has also been obtained 
analytically in Ref.\,\cite{KMP} in the region $L \alt N$ with use of 
a Bogoliubov transformation, and it is also in agreement with the study of 
this problem using the mean-field approximation \cite{Rokhsar,KMP}.

Here, we present an analytic derivation of this linear dependence of the 
ground state energy on $L$.  Our starting point is the hamiltonian 
$\hat H=\hat H_0 + \hat V$, where
\begin{eqnarray}
    \hat H_0 = \sum_{i} - \frac {\hbar^{2}} {2M} {\ve \nabla}_{i}^{2} +
  \sum_i  \frac 1 2 \, M \omega^{2} r_{i}^{2}
\label{h0}
\end{eqnarray}
includes the kinetic energy of the particles and the potential energy
due to the trapping potential, and  
\begin{eqnarray}
   \hat V = \frac{1}{2} U_{0} \sum_{i \neq j} \delta({\bf r}_{i} - {\bf r}_{j})
\label{v}
\end{eqnarray}
is the interaction energy, which we assume to be a contact interaction.
Here, $M$ is the mass of the atoms, $\omega$ is the frequency of the trapping 
potential, which is assumed to be isotropic, and $U_0 = 4 \pi \hbar^2 a/M$ 
is the strength of the effective two-body interaction, with $a$ being the 
scattering length for atom-atom collisions.  We assume that $a > 0$, i.e., 
we treat only the case of repulsive interactions between the atoms.  Further, 
we restrict our attention to the domain 
\begin{eqnarray}
   n U_0 \ll \hbar \omega,
\label{wi}
\end{eqnarray}
where $n$ is the density of the condensed atoms.  If the system has angular 
momentum $L$, this condition allows us to work within the subspace of states 
which are degenerate in the absence of interactions.  All other states
differ by an energy of order $\hbar \omega$, which is much larger than 
$n U_0$ by assumption. 

It is convenient to work in the basis 
\begin{equation}
  |N_0,N_1,N_2,... \rangle \equiv
 |0^{N_0}, 1^{N_1}, 2^{N_2}, ...\rangle,
\label{states}
\end{equation}
where $N_k$ is the number of particles with angular momentum $\hbar k$.
These occupation numbers must be chosen to respect the obvious constraints 
on particle number and angular momentum:
\begin{eqnarray}
   \sum_k N_k = N; \ \ \sum_k k N_k = L.
\label{restr}
\end{eqnarray}
We introduce annihilation and creation operators $a_k$ and $a_k^\dagger$,
which destroy or create one particle with angular momentum $\hbar k$.  The 
number operator $\hat N$ and the angular-momentum operator $\hat L$ can 
then be expressed as
\begin{eqnarray}
  \hat N = \sum_k a_k^\dagger a_k; \, \hat L = \sum_k \hbar k a_k^\dagger 
a_k .
\label{noper}
\end{eqnarray}
We also introduce the raising and lowering operators, ${\hat L}_{\pm}$, as  
\begin{eqnarray}
  \hat L_+ = \sum_k \sqrt{k+1} \, a_{k+1}^\dagger a_k; \,
 \hat L_- = \sum_k \sqrt{k+1} \, a_k^\dagger a_{k+1}.
\label{lrop}
\end{eqnarray}
Finally, we write the interaction $\hat V$, Eq.\,(\ref{v}), as 
\begin{eqnarray}
	 \hat V = \frac {U_0} 2 \sum_K \ \left( \sum_{k'=0}^K \
\frac{a_{k'}^{\dag}}{\sqrt{I_{k'}}} \frac{a_{K-k'}^{\dag}}{\sqrt{I_{K-k'}}}
\right) J_K 
\nonumber \\
\times \left( \sum_{k=0}^K \
\frac{a_{k}}{\sqrt{I_{k}}} \frac{a_{K-k}}{\sqrt{I_{K-k}}}
\right) \ ,
\label{interd}
\end{eqnarray}
where $I_k = k!$ and $J_K = {K!}/{2^K}$.  It is readily seen that 
\begin{eqnarray}
 [\hat L, \hat L_{\pm}] = \hat L_{\pm},
\label{commut}
\end{eqnarray}
which shows that the operators $\hat L_{\pm}$ raise or lower the angular 
momentum by one unit when applied to the states of Eq.\,(\ref{states}). 
Further, both the interaction, $\hat V$, and the hamiltonian commute with 
$\hat L_{\pm}$,
\begin{eqnarray}
  [{\hat V}, {\hat L}_{\pm}] = [{\hat H}, {\hat L}_{\pm} ] = 0.
\label{commutlpm}
\end{eqnarray}
This shows that, if $| L,N \rangle$ is an eigenstate of ${\hat V}$ with 
eigenvalue ${\cal E}_{L,N}$, the state ${\hat L}_+ |L,N\rangle$ will also 
be an eigenstate of ${\hat V}$ with the same eigenenergy but with $L+1$ 
units of angular momentum.  

We now proceed to the demonstration that the spectrum is indeed linear
if $L \le N$.   For any given eigenstate with a given $N$ and $L$, the 
repeated application of ${\hat L}_{+}$ will generate infinitely many new 
eigenstates with angular momenta $L+1, L+2, ...$.  These states are 
obtained by adding angular momentum to the center of mass coordinate 
\cite{Wilkin,Ben,Kivelson} while leaving the wavefunction unchanged in the 
relative coordinates of the particles.  Since the interaction energy depends 
only on relative coordinates and since the center of mass variables separate 
for a harmonic potential, the interaction energy is the same for each 
of these states.  For any given $N$ and $L$, it is useful to pay special 
attention to those ``intrinsic'' eigenstates, $| L, N \rangle_{\rm int}$, 
which do not have any center of mass excitation.  The observation that the 
interaction energy in the ground state decreases with increasing $L$ 
\cite{Bertsch} indicates that the ground state wavefunction is an 
intrinsic state.  When ${\hat L}_+$ acts on any eigenstate $|L-1,N \rangle$ 
(i.e., intrinsic or not), it creates a state $|L,N \rangle$ with center of 
mass excitation.  Such states are necessarily orthogonal to intrinsic 
states with angular momentum $L$, and thus
\begin{eqnarray}
   _{\rm int}\langle L, N | \hat L_+ | L-1,N \rangle = 
 \langle L-1, N | \hat L_- | L, N \rangle_{\rm int} = 0.
\label{orth}
\end{eqnarray}
Since the states $| L-1, N \rangle$ form a complete set with angular 
moment $L-1$, we conclude from Eq.\,(\ref{orth}) that 
\begin{eqnarray}
  \hat L_- | L, N \rangle_{\rm int} = 0.
\label{van}
\end{eqnarray}

Let us now consider two particular basis states, $|A\rangle$ and 
$|B\rangle$, for any $L < N$ in the occupation number representation 
of Eq.\,(\ref{states}),
\begin{eqnarray}
 |A \rangle = |N-L+1, L-2, 1 \rangle;\ |B \rangle = |N-L, L, 0 \rangle.
\label{AB}
\end{eqnarray}
States $|A\rangle$ and $|B\rangle$ are not eigenstates of ${\hat V}$ but 
will in general be components of these eigenstates.  Similarly, consider 
the state $|C\rangle$
\begin{eqnarray}
 |C \rangle = |N-L+1, L-1, 0 \rangle.
\label{C}
\end{eqnarray}
Note that these three states have $N_k = 0$ for all $k \ge 3$.  Under the 
action of ${\hat L}_{-}$, $|A\rangle$ and $|B\rangle$ are the only states 
which can lead to state $|C\rangle$.  Therefore, if an intrinsic eigenstate 
$|L, N \rangle_{\rm int}$ includes these basis states in the combination 
$\alpha |A \rangle + \beta |B \rangle$, we see from Eq.\,(\ref{van}) that  
\begin{eqnarray}
   \hat L_- ( \alpha |A \rangle + \beta |B \rangle ) \propto |C\rangle = 0.
\label{exp}
\end{eqnarray}
We can satisfy Eq.\,(\ref{exp}) either by choosing $\alpha = \beta = 0$ or 
by insisting that
\begin{eqnarray}
  \frac {\beta} {\alpha} = - \left( \frac {{2 (L-1)}} 
 {L (N-L+1)} \right)^{1/2}.
\label{condition}
\end{eqnarray}
The second option is interesting since it uniquely determines the ratio 
of two components in $|L, N \rangle_{\rm int}$.

Since the state $|L, N \rangle_{\rm int}$ is an eigenstate of ${\hat V}$, 
it follows that  
\begin{eqnarray}
  \langle B | \hat V | L, N \rangle_{\rm int} = {\cal E}_{L,N}
 \langle B | L, N \rangle_{\rm int}.
\label{me}
\end{eqnarray}
However, the only component of the state $|L, N \rangle_{\rm int}$ which
can be connected to state $|B \rangle$ by the interaction
$\hat V$ is precisely the linear superposition $\alpha |A \rangle +
\beta |B \rangle$.  Since the ratio $\beta / \alpha$ is known, this 
fact can be used to determine the eigenvalue ${\cal E}_{L,N}$.  Since 
$|A\rangle$ and $|B \rangle$ involve particles with $k=0,\ 1$, and 2 only, 
the relevant part of the interaction $\hat V$ is thus 
\begin{eqnarray}
   \hat V = \frac 1 2 U_0 ( a_0^\dagger a_0^\dagger a_0 a_0
   + 2 a_1^\dagger a_0^\dagger a_1 a_0
  + \frac 1 2 a_1^\dagger a_1^\dagger a_1 a_1 
\nonumber \\ 
 + \frac 1 {\sqrt 2} a_1^\dagger a_1^\dagger a_2 a_0
+ \frac 1 {\sqrt 2} a_2^\dagger a_0 ^\dagger a_1 a_1 ). 
\label{v012}
\end{eqnarray}
The required matrix elements are 
\begin{eqnarray}
  \frac \beta 2 \langle B | \, a_0^\dagger a_0^\dagger a_0 a_0 + 
 2 a_1^\dagger a_0^\dagger a_1 a_0 + \frac 1 2 
a_1^\dagger a_1^\dagger a_1 a_1 | B \rangle =
\nonumber \\
  \frac \beta 2 [(N-L)(N-L-1) + 2 L (N-L) + \frac 1 2 L (L-1)],
\label{me1}
\end{eqnarray}
and
\begin{eqnarray}
  \frac \alpha {2 \sqrt 2} \langle B | \, a_1^\dagger a_1^\dagger a_2 a_0 
 | A \rangle = \frac \alpha {2 \sqrt 2} \sqrt{(N-L-1)(L-1) L}.
\label{me2}
\end{eqnarray}
It now follows directly from Eqs.\,(\ref{condition}), (\ref{me}), 
(\ref{me1}), and (\ref{me2}) that
\begin{eqnarray}
   {\cal E}_{L,N} = \frac {N(2N-L-2)} 4 v_0,
\label{energy}
\end{eqnarray}
in agreement with the numerical results of Refs.\,\cite{Bertsch,KMP}.  
Here $v_0 = U_0 \int |\Phi_0({\bf r})|^4 \,d{\bf r}$, and $\Phi_0$ is 
the ground state wavefunction of the harmonic oscillator. To be 
precise, we have demonstrated that an intrinsic eigenstate {\em either\/} 
has the eigenvalue Eq.\,(\ref{energy}) {\em or\/} does not contain the 
components $| A \rangle$ and $| B \rangle$. Completeness indicates that
at least one intrinsic eigenstate with eigenvalue given by 
Eq.\,(\ref{energy}) must exist. 

  As we mentioned earlier, the interaction energy of Eq.\,(\ref{energy})
varies linearly with $L$ for the contact two-body potential
of Eq.\,(\ref{v}). This feature is more general.
As long as one deals with a two-body potential that conserves the angular
momentum and commutes with $\hat L_\pm$, this linearity persists, with the
slope $\partial {\cal E}_{L,N}/\partial L$ being dependent on the specific form
of the interaction. 

We have exploited the special features of harmonically trapped, weakly 
interacting bosons to determine a single eigenvalue of ${\hat V}$ from 
the properties of two (in general exponentially small) components of the 
wavefunction.  The present elementary argument does not enable us to 
construct the full eigenfunction or to demonstrate that this eigenstate 
describes the ground state of the system.  While the present method can be 
applied to other choices of ${\hat V}$, there appears to be no simple 
generalization to the interesting case $L > N$.   

\vskip1pc

We thank B. Mottelson, C. J. Pethick and S. M. Reimann for helpful 
discussions.  G.M.K. was supported by the European Commission, TMR program, 
contract No.\,ERBFMBICT 983142.  G.M.K. would like to thank the Foundation of 
Research and Technology, Hellas (FORTH) for its hospitality.

\end{document}